\documentclass[aps,a4paper,twocolumn,superscriptaddress]{revtex4-1}
\pdfoutput=1
\usepackage[]{latexsym,amsmath,amssymb,graphicx,epsfig}

\newcommand{\be}{\begin{eqnarray}}
\newcommand{\ee}{\end{eqnarray}}
\newcommand{\degree}{\ensuremath{^\circ}}

\begin{document}


\title{Quantum Black Holes Effects on the Shape of Extensive Air Showers}



\author{Nicusor Arsene}
\email[]{nicusorarsene@spacescience.ro}
\affiliation{Institute of Space Science, P.O.Box MG-23, Ro 077125 Bucharest-Magurele, Romania}
\affiliation{Physics Department, University of Bucharest, Bucharest-Magurele, Romania}

\author{Lauretiu Ioan Caramete}
\email[]{lcaramete@spacescience.ro}
\affiliation{Institute of Space Science, P.O.Box MG-23, Ro 077125 Bucharest-Magurele, Romania}

\author{Peter B. Denton}
\email[]{peterbd1@gmail.com}
\affiliation{Vanderbilt University, Nashville, TN 37235}

\author{Octavian Micu}
\email[]{octavian.micu@spacescience.ro}
\affiliation{Institute of Space Science, P.O.Box MG-23, Ro 077125 Bucharest-Magurele, Romania}


\date{\today}

\begin{abstract}
We investigate the possibility to find a characteristic TeV scale quantum black holes decay signature in the data recorded by cosmic rays experiments. TeV black holes can be produced via the collisions of ultra high energetic protons $(E>10^{18}eV)$ with nucleons from the from atmosphere. We focus on the case when the black hole decays specifically into two particles. These particles are then boosted in the Earth reference frame (back-to-back in the center of mass reference frame) and induce two overlapping showers.
When reconstructing both the energy and the shape of the resultant air shower, there is a significant difference between showers induced only via standard model interactions and showers produced via the back-to-back decay of black holes as intermediate states.
\end{abstract}

\pacs{}

\maketitle

\section{Introduction}

Brane world models \cite{ArkaniHamed:1998rs, Antoniadis:1998ig, Randall:1999ee} or even four dimensional models with a large hidden sector of particles \cite{Calmet:2008tn} have been suggested when trying to explain the large hierarchy between the strength of the gravitational force and the standard model. In this context quantum gravity can become important anywhere between the standard Planck scale (i.e. some $10^{16}$ TeV) and a few TeV. When the energy scale of gravity is in the lower end of this energy range (energies  accessible for particle accelerators or in the center of mass of the collisions between ultra high energy cosmic rays and nucleons from the atmosphere) particle collisions can result in the creation of TeV mass black holes. This is a threshold effect in the sense that black hole creation turns on when the center of mass energy reaches the Planck scale. 

Black holes formation via particle collisions has been studied since  the 70's. The {\em Hoop conjecture\/} proposed by K.~Thorne in 1972 ~\cite{Thorne:1972ji} states that  a black hole forms whenever the impact parameter $b$ of two colliding objects (of negligible spatial extension) is shorter than the radius of the would-be-horizon (roughly, the Schwarzschild radius, if angular momentum can be neglected)
corresponding to the total energy $M$ of the system \footnote{We
shall use units with $c=\hbar=1$ and the Boltzmann constant $k_{B}=1$,
and always display the Newton constant $G=l_{Pl}/M_{Pl}$, where $l_{Pl}$ and $M_{Pl}$
are the Planck length and mass, respectively.}
\begin{eqnarray}
b\lesssim \frac{2\,l_{Pl}\,M}{M_{Pl}}
\ .
\label{hoop}
\end{eqnarray}
This is intuitive but not enough to prove that black holes do indeed form in such collisions. However, there are now proofs (the first
ones performed by Penrose who never published his findings) for the formation of closed trapped surfaces, which are enough to
demonstrate gravitational collapse and hence black hole formation. Refs.~\cite{D'Eath:1992hb, D'Eath:1992hd, D'Eath:1992qu,
Eardley:2002re} cover both the cases of zero and non-zero impact parameters. The analytical proof of Eardley and Giddings for the case
of a four dimensional space-time \cite{Eardley:2002re} demonstrates the formation of classical black holes due to the collisions of two
particles with a non-zero impact parameter at energies much larger than the Planck mass. The proof was extended to the semi-classical
regime (semi-classical black holes are objects with masses in the range from 5 to 20 times the Planck mass \cite{Meade:2007sz}) by Hsu \cite{Hsu:2002bd}. Black hole formation via the collision of two quantum mechanical wave packets for the case of zero impact parameter was also recently shown in ref. \cite{Casadio:2013uga}. 

Many articles have considered semi-classical TeV mass black holes production at particle colliders or in the cosmic ray data
\cite{Dimopoulos:2001hw, Banks:1999gd, Giddings:2001bu, Feng:2001ib, Anchordoqui:2003ug, Anchordoqui:2001cg, Anchordoqui:2003jr,
Kowalski:2002gb, Ringwald:2001vk}. The possibility also exists for the energy in the center of mass not to be large enough for
semi-classical black holes to be produced and it was proposed \cite{Calmet:2008dg, Calmet:2011ta, Calmet:2012cn} to also consider
quantum black holes. These are non-thermal objects with masses up to five Planck masses which are also easier to produce. Because they
are non-thermal, quantum black holes are expected to decay into a small number of particles, typically two. Experimental signatures for
such decays are very different from the one of semi-classical objects which are expected to decay into several particles in a final
explosion, see e.g.~\cite{Cavaglia:2002si, Kanti:2004nr} for recent reviews. 

Refs.~\cite{Calmet:2012mf, Arsene:2013nca} investigate the possibility to detect the back-to-back decays of TeV scale quantum black
holes by observing double shower events (showers having common origins and developing at an angle) in the cosmic ray data or similar
events in the data recorded by neutrino observatories. In the latter case one would observe muon tracks starting from a common origin
and oriented at an angle.
Such black holes are produced in the collisions between protons or neutrinos with energies above $10^{17}$ eV and nucleons from the
atmosphere respectively water or ice. The black holes immediately decay into two standard model particles. The decays for which two
distinct showers are visible represent less than one percent from the total number of black hole events.
Ref.~\cite{Luo:2013iia} discusses experiments and simulations of the presence of multi-core showers.
In 99.9$\%$ of the cases the two showers overlap entirely. 

It needs to be pointed out that this signature, along with the ones proposed in Refs.~\cite{Calmet:2012mf, Arsene:2013nca} are
complementary to the TeV scale gravity searches performed by the various experimental groups from the Large Hadron Collider (LHC)
\cite{CMS:2012yf, ATLAS:2012pu}. As it will become obvious later, the signature proposed here actually allows the community to look for
the scale of gravity in the tens of TeV regime, energies beyond the reach of any current particle physics experiment.    

In this article we study the possibility to distinguish the extensive air showers induced by back-to-back black hole decays
from standard showers. Experiments such as Pierre Auger Observatory \cite{Abraham:2004dt} and Telescope Array \cite{Tokuno:2012mi} can
evaluate the shape of showers with their fluorescence detectors and the energy deposited by the shower in surface detectors. Space based experiments such as the
proposed space based JEM-EUSO experiment \cite{TheJEM-EUSO:2013vea} will provide an additional means to detect the shape of showers at energies
above $10^{19}$ eV. The fluorescence detectors are used to determine the mass composition of primary particle by measuring the
atmospheric depth where the density of charged particles is maximum (so called $X_{max}$) and at the same time to estimate the energy
of the primary particle by integrating the Gaisser-Hillas curve \cite{Gaisser:111} and multiplying by a mean energy loss rate in the
atmosphere of $2.19$ MeV/g\;cm$^{-2}$. For the ground based experiments, the energy of the primary particle can also be calculated using the energy deposited in 
the grid of surface detectors which consists in 1600 water Cherenkov tanks placed at a distance of  1.5 km each other, and are dedicated to measure the lateral distribution function (LDF) of the showers. Using the signal recorded by detectors situated 1000 meters away from the shower axis, S(1000), one can estimate the energy of primary particle.   

The findings of the present article are based on a set of extensive air shower simulations made using CORSIKA 6.990 (COsmic Ray SImulations for KAscade) \cite{corsika, corsika1} for micro black holes produced by protons with energies of $10^{18}$ eV which interact with nuclei in the atmosphere. The black holes decay immediately back-to-back into two particles, in our case a pair of pions. Two possibilities will be considered, quantum black holes decaying into pairs of $\pi^{+}$ and $\pi^{-}$ ($QBH\rightarrow\pi^{+}\pi^{-}$) and quantum black holes decaying into two $\pi^{0}$ mesons ($QBH\rightarrow\pi^{0}\pi^{0}$). The two black hole decay products then produce overlapping extensive atmospheric showers. 
As it is well known, gravity is democratic and black holes decay into any elementary standard model particles. The reason for choosing two hadrons is that, when compared with decays into charged leptons or photons, the $X_{max}$ values are always smaller for showers induced by hadrons of identical energies. Moreover, it will be seen that even in the case of neutral pions which decay predominantly into photons, the $X_{max}$ values will increase comparatively to the ones of charged pions.
The numerical simulations will focus on the case in which the particles have approximately equal energies in the laboratory reference frame.  As it will become obvious from the numerical simulations, the air showers look very different for the two decay channels considered.

\section {Black holes production} 

The number of black holes expected to be produced within the volume of the atmosphere visible to a cosmic rays experiment, taking into account the experiment's dimensions and the duty cycle of the detectors, is given by
 \begin{eqnarray}
N&=&  \int dE N_A  \frac{d\Phi}{dE} \sigma(E) A(E) T
\label{N}
\end{eqnarray} 
where  $\sigma(E)$ is the production cross section described bellow, $\frac{d\Phi}{dE}$ is the flux of cosmic ray particles, $A(E)$ is the acceptance of the experiment measured in cm$^2$ sr yr, $N_A$ is Avogadro's number and $T$ is the running time of the detectors.

The cross section p N $\to$ BH is given by:
\begin{eqnarray}
\sigma^{pN}(s,x_{min},n,M_D) &=& \int_0^1 2z dz \int_{\frac{(x_{min} M_D)^2}{y(z)^2 s}}^1 du \\ \nonumber && \times
 \int_u^1 \frac{dv}{v}  F(n) \pi r_s^2(us,n,M_D)
\\
\nonumber  &&\times
 \sum_{i,j} f_i(v,Q) f^N_j(u/v,Q)
\end{eqnarray}
where $M_D$ is the $4+n$ dimensional reduced Planck mass, $z=b/b_{max}$, $x_{min}=M_{BH,min}/M_D$,  $n$ is the number of extra-dimensions, $F(n)$ and $y(z)$ are the factors introduced by Eardley and Giddings \cite{Eardley:2002re} and by Yoshino and Nambu \cite{Yoshino:2002br}. The virtuality scale $Q$ is taken to be of the order of $M_D$. The $4 + n$ dimensional  Schwarzschild radius is given by 
\begin{eqnarray}
r_s(us,n,M_D)=k(n)M_D^{-1}[\sqrt{us}/M_D]^{1/(1+n)}
\end{eqnarray}
where
\begin{eqnarray}
k(n) =  \left [2^n \sqrt{\pi}^{n-3} \frac{\Gamma((3+n)/2)}{2+n} \right ]^{1/(1+n)}.
\end{eqnarray}

Note that $s= 2 x m_N E$, with $m_N$ the  nuclei mass and $E$ the cosmic ray energy. The functions $f_i(x,Q)$ are the parton distribution functions.  Black holes can also be produced from neutrinos interacting with nucleons in the atmosphere, but this production rate might be suppressed in comparison to the production rate from UHECRs \cite{Stojkovic:2005fx}.

The number of the black holes depends directly on the flux of the cosmic ray particles. It is important to note that the composition of
the cosmic ray flux includes neutrons, protons, neutrinos, light, intermediate and heavier nuclei like Fe \cite{Allard:2008gj}. Since it seems that protons are the main constituent particles in the cosmic ray flux at $10^{18}$ eV, we select protons as primary particles to perform our analysis and comparisons (it was shown in \cite{Stojkovic:2005fx} that the production rate of black holes from neutrinos is suppressed).

\section{Black holes decay}

We wish to analyze the signature generated by the two overlapping showers induced by the decay products of a quantum black hole.
This is an extension of the cases studied previously in \cite{Calmet:2012mf, Arsene:2013nca}. More specifically, in the previous papers
the authors analyzed the possibility for the particles resulting from the back-to-back decay of quantum black holes to generate showers
which are separated spatially. As it turned out, only for less than 1\% of the quantum black hole decays are the two showers separated
spatially. In this article we analyze the signature of the events in which quantum black holes decay into a pair of pions (which could
be any of the $\pi^{0}$, $\pi^{+}$, $\pi^{-}$ depending on the intermediary quantum black hole electric charge), particles which then induce two
overlapping showers. 
One also has to take into account the possibility that the black holes decay into two partons such as quarks which would hadronize after traveling over distances of some 200$^{-1}$ MeV and become an SU(3)$_c$ singlets. The creation of two pions during the hadronization process is very likely since mesons consist of only quark-antiquark pairs.
We emphasise again, that because of the heavy simulations involved, we only focus on the case in which the two pions have roughly equal energies in the Earth reference frame.   

The process of black hole formation requires for the impact parameter $b$ (defined as the perpendicular distance between the paths of the two particles that are colliding) to be smaller than the horizon radius and we will only consider the events for which this inequality holds. Also in the process of black hole formation via particle collisions, some energy is radiated as gravitational radiation. We will work with a further simplifying assumption, which is that the whole energy of the two particles, including the partons of the protons goes into the black hole creation. Using a simple relativistic textbook calculation \cite{Calmet:2012mf, Arsene:2013nca} one can calculate the black hole mass $M_{BH}$ and relativistic Lorentz factor $\gamma_{BH}$. 

 As stated before, quantum black holes are non-thermal objects which decay into a small number of particles, most likely into two particles moving back-to-back in the center of mass reference frame and with no preferred direction with respect to the direction of motion of the black hole. The main constraints on the decay are for the sum of the masses of the two resulting particles to be smaller than the black hole mass $M_{BH}$ and for the standard model charges to be conserved. 

\begin{figure}[t]
\includegraphics[scale=0.45]{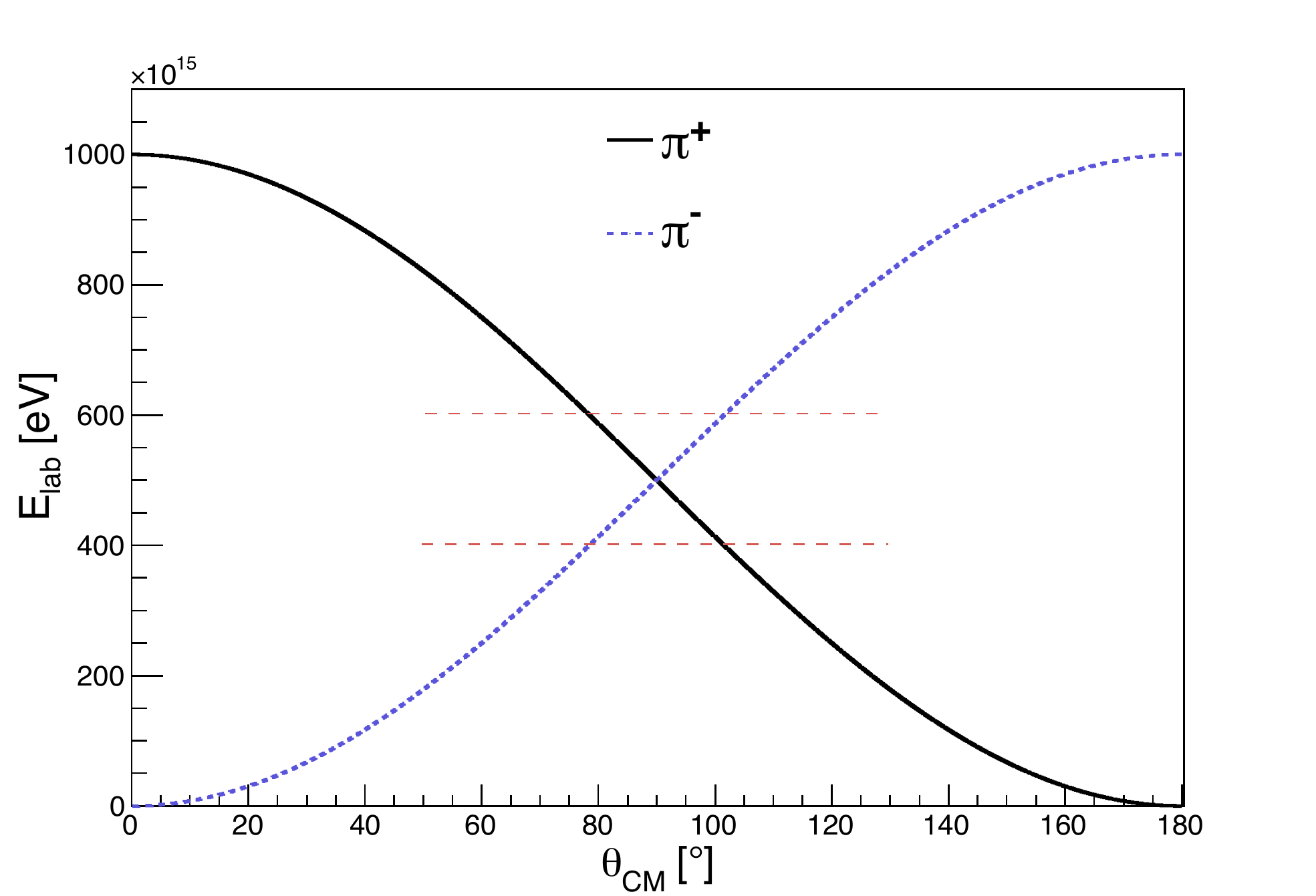}
\caption{The energies of the two resulting particles (for our simulations we consider a $\pi^{+}$ and $\pi^{-}$, but the same is true for the case of two neutral pions since their masses are roughly the same) in the Earth reference frame a a function of the decay angles measured from the direction of propagation of the quantum black hole in the center of mass frame.  The red dotted lines highlight the intervals of angles for which the energies of the two particles vary between $4\times10^{17} - 6\times10^{17}$ eV.} 
\label{plotEnergiesVsAngles}
\end{figure} 

Under the assumptions stated above, when a proton having an energy of $10^{18}$ eV collides with a nucleon in the atmosphere, the
resulting quantum black hole mass can be up to the order of $M_{BH} \simeq 4 \times 10^{13}$ eV and can move relativistically
with a gamma factor of $\gamma_{BH} \simeq 2\times 10^4$. Such large gamma factors have significant impact on the angle between the
trajectories of the two particles when viewed from the Earth reference frame. Also the energies of the two particles, when measured in
this reference frame, vary due to a combination of the Lorentz factor of the center of mass and the directions at which the two
particles move in the center of mass reference frame with respect to the direction of motion of the center of mass. These dependencies
are encoded in the Lorentz transformation formulas and for the particular case discussed here this dependency is shown in
Fig.~\ref{plotEnergiesVsAngles}. The plot represents the energy in the laboratory/Earth reference frame as a 
function of the angle that the trajectories of the particles make in the center of mass measured with respect to the direction of motion of the center of mass. Because of limited computational power (one simulation requires one processor core to run at full power on the order of a week) we limit our simulations to the case in which the energies of the two particles (in the Earth reference frame) are roughly equal. Therefore the present analysis will apply to those cases. One might extend the simulations for energies which vary on a broader range (one with respect to the other). Therefore, for our case of interest we select the interval of angles for which the two pions have energies between $4\times10^{17} - 6\times10^{17}$ eV. This happens for $75\degree\leq \theta_{CM} \leq 105\degree$. One can easily calculate that for $25.8\%$ of the total number of quantum black holes produced the particles resulting from their back-to-back decay are emitted in this interval of angles.

Using this range of angles, together with the acceptance for the Pierre Auger Observatory 
\cite{2011arXiv1107.4809T} and a fit for the cosmic ray flux \cite{2008PhRvL.101f1101A} in Eq.~\ref{N} one can estimate the number of
events of this type that are expected to be seen in the Pierre Auger Observatory data. Another ingredient needed in Eq.~\ref{N} is the
extra-dimensional scenario considered and we will analyze the cases for $n= 0, 1, 4, 5, 6, 7$ extra dimensions,  where the case $n= 0$
refers to a scenario with no extra-dimensions but in which low scale gravity is due to the existence of a large hidden sector of
particles which interact only gravitationally \cite{Calmet:2008tn}. 
Also the case $n=1$ corresponds to the Randall-Sundrum model since the ADD scenario with one extra-dimension is already excluded by other experiments. 

The number of quantum black holes which can be created also depends on the value of the Planck scale. As stated before we are interested in quantum black holes which have masses between one and five Planck masses. Considering that a  $10^{18}$ eV cosmic ray produces a black hole on the order of 40 TeV, this is a quantum black hole only if the Planck scale is around $5$ TeV or greater. This means that this signature can be used to search for the possibility that the Planck scale is above $5$ TeV and so a natural extension of the LHC searches. Table \ref{bhNumbers} shows the number of quantum black holes for which the energies of the two particles they decay into are between $4\times10^{17} - 6\times10^{17}$ eV when measured in the Earth reference frame as function of the number of extra-dimensions and the value of the Planck mass. 


\begin{widetext}

\begin{table}
\begin{tabular}{||c|c|c|c|c|c|c||}
\hline\hline
No. of extra dimensions &  $M_{Pl}=5$ TeV &  $M_{Pl}=6$ TeV & $M_{Pl}=7$ TeV &  $M_{Pl}=8$ TeV  &  $M_{Pl}=9$ TeV & $M_{Pl}=10$ TeV \\ \hline\hline
0&29&14&7.5&4.5&2.8&1.8\\ \hline
1&1.9$\times 10^2$&1.1$\times 10^2$&0.71$\times 10^2$&0.48$\times 10^2$&0.34$\times 10^2$&0.25$\times 10^2$\\ \hline
4&1.1$\times 10^3$&7.6$\times 10^2$&5.2$\times 10^2$&3.8$\times 10^2$&2.9 $\times 10^2$&2.2$\times 10^2$\\ \hline
5&1.6$\times 10^3$&1.0$\times 10^3$&7.0$\times 10^2$&5.2$\times 10^2$&4.0$\times 10^2$&3.1$\times 10^2$\\ \hline
6&2.0$\times 10^3$&1.3$\times 10^3$&9.2$\times 10^2$&6.7$\times 10^2$&5.2$\times 10^2$&4.0$\times 10^2$\\ \hline
7&2.4$\times 10^3$&1.6$\times 10^3$&1.1$\times 10^3$&8.3$\times 10^2$&6.4$\times 10^2$&5.0$\times 10^2$\\ \hline
\hline
\end{tabular}
\caption{Number of black hole events per year expected at the Pierre Auger Observatory experiment for which the angle between the direction of the two decaying particles and the direction of motion of the quantum black hole lies in the interval between $75\degree - 105\degree$ in the center of mass frame.}
\label{bhNumbers}
\end{table}

\end{widetext}

\section {Simulations and Results}

CORSIKA is a code based on Monte Carlo methods, dedicated to simulate in detail the development of extensive air showers in the atmosphere. For our simulations we chose the altitude, observation plane and magnetic field for the position of the Pierre Auger Observatory. CORSIKA also allows performing cuts for the energies of the particles. For the simulations performed energy cuts of the secondary particles are set at 300 MeV for hadrons and for muons; and 3 MeV for electromagnetic component. For the simulations we use the QGSJET 01C model \cite{1997NuPhS..52...17K} for high energy hadronic interactions.

\begin{figure}[ht]
\includegraphics[scale=0.59]{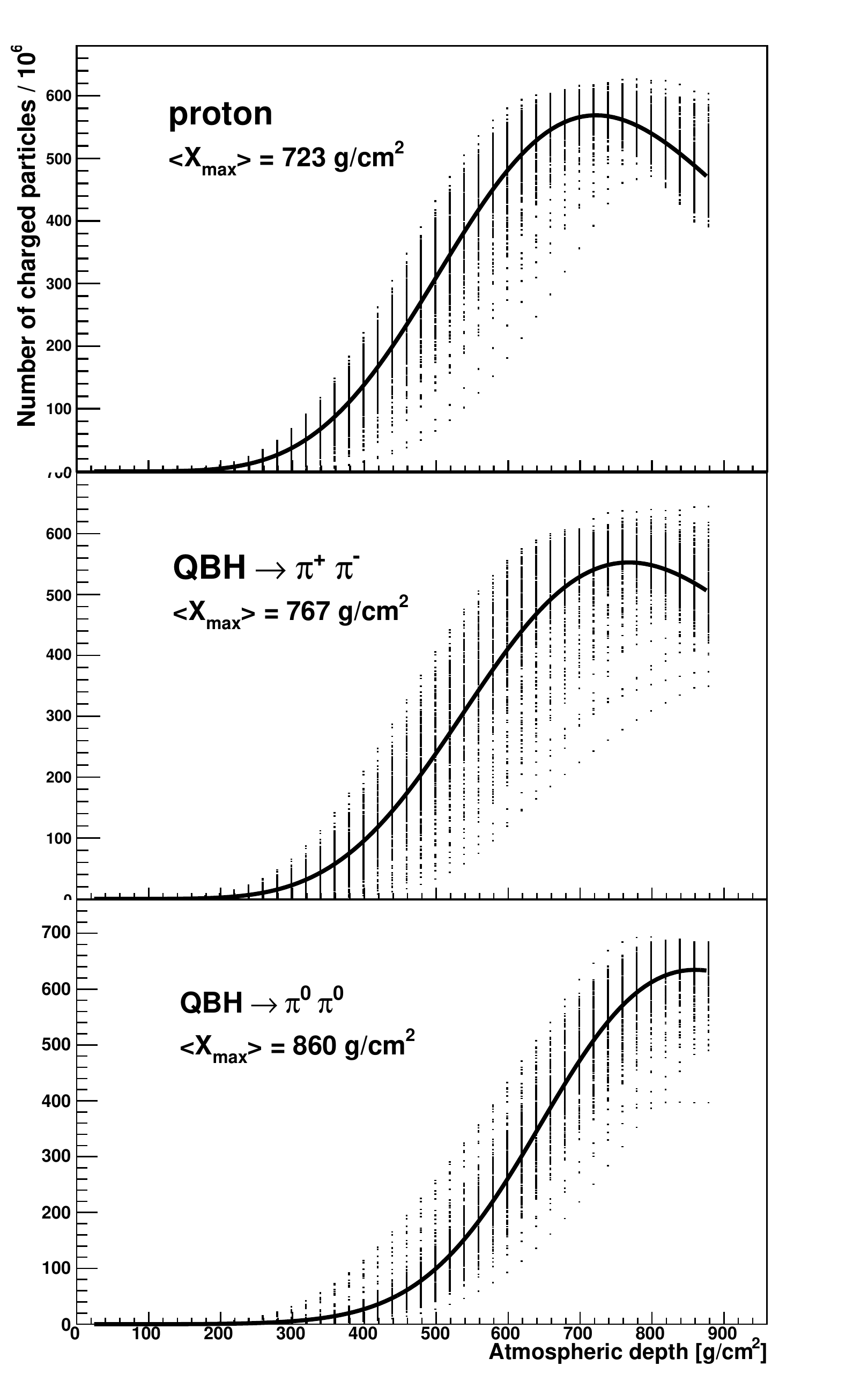}
\caption{Longitudinal profiles of extensive air showers simulated with CORSIKA. Each plot contains 250 simulations. The plot on top shows standard model showers induced by protons of energies $E=10^{18}$ eV. The two lower plots contain black hole induced shower simulations. In these cases
the primary particles were $E=10^{18}$ eV protons which produced quantum black holes that decayed into $\pi^{+}\pi^{-}$ (middle), respectively $\pi^{0}\pi^{0}$ (bottom), with roughly equal
energies in the laboratory reference frame. The black dots represent the number of charged particles for each simulation and the black lines
represent the fits with the Gaisser-Hillas function to obtain the mean values of $X_{max}$.}
\label{plot1}
\end{figure} 

Further, we wish to analyze if there is a distinctive signature for an extensive air shower produced via the back-to-back decay into two particles of a black hole when compared with a standard air shower produced by protons. In each of the cases the starting particles are protons with energies of $10^{18}$ eV. For what we call standard air showers the protons interact with nuclei from the atmosphere and produce the usual showers which are recorded by cosmic ray observatories. We call "black hole induced showers" the showers for which protons first interact with nucleons to create quantum black holes. The black holes decay instantaneously back to back into two particles which are highly boosted forward in the Earth reference frame. We consider two possible decay channels: quantum black holes decaying into pairs of $\pi^{+}$ and $\pi^{-}$ ($QBH\rightarrow\pi^{+}\pi^{-}$) and quantum black holes decaying into two $\pi^{0}$ mesons ($QBH\rightarrow\pi^{0}\pi^{0}$). 
For the numerical simulations that we perform, we consider the case when the two pions have roughly equal energies (on the order of $5 \times 10^{17}$ eV) in the Earth reference frame. These pions further interact with nucleons to produce extensive air showers. In the following paragraphs we make a thorough comparison between the standard proton induced showers and the black hole induced showers. The two different quantum black hole decay channels are considered separately. The primary interaction point is taken at 20 km altitude in both types of simulations. This is the average altitude at which protons with this energy moving vertically first interact in the atmosphere. All the results of the simulations will be presented in units of atmospheric depth in any case. 

\begin{figure}[ht]
\includegraphics[scale=0.47]{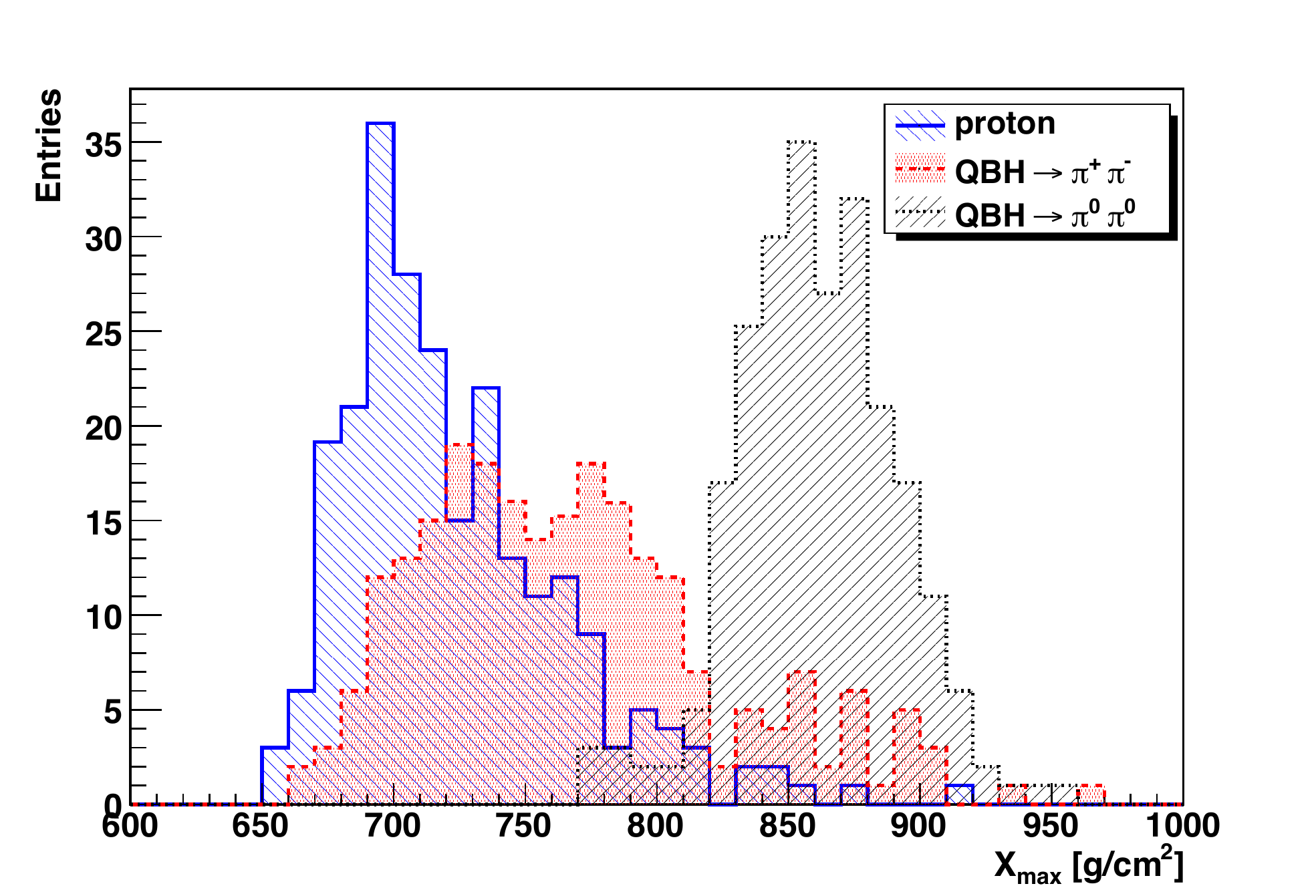}
\caption{{Distributions of all the individual $X_{max}$ values for the simulations included in Fig.~\ref{plot1}.}} 
\label{plot1.1}
\end{figure} 

We performed a set of 250 CORSIKA simulations for each case: the standard model type of proton induced showers and the two quantum black hole decay channels. We found the following values: $<X_{max}^{p}> =$ 723 g/cm$^{2}$ with the RMS = 42 g/cm$^{2}$ (in very good agreement with the QGSJET01C model), $<X_{max}^{QBH\rightarrow\pi^{+}\pi^{-}}>$ = 767 g/cm$^{2}$ with the RMS = 50 g/cm$^{2}$ and $<X_{max}^{QBH\rightarrow\pi^{0}\pi^{0}}>$ = 860 g/cm$^{2}$ with the RMS = 30 g/cm$^{2}$. Fig.~\ref{plot1} shows a comparison between the $<X_{max}>$ values for these three cases. A shift of approximately 44 g/cm$^{2}$ is observed for the case of the showers induced by black holes which decay into a pair of charged pions when compared with standard model showers. An even larger shift of 137 g/cm$^{2}$ is observed in the case when the black holes decay into a pair of neutral pions. One can see that the $<X_{max}>$ values for black hole induced showers are the largest ones. These differences (specially the second one) are significant, considering that the difference in $<X_{max}>$ between a proton shower and iron nucleus shower at the same energy is $<X_{max}^{p}> - <X_{max}^{Fe}> \simeq 100$ g/cm$^{2}$. This difference can be observed by the fluorescence detectors of the cosmic ray observatories. The distribution of the individual $X_{max}$ values for all of the simulations included in Fig.~\ref{plot1} is presented in Fig.~\ref{plot1.1}. We emphasize once more so far we fixed the height of primary interaction at 20 km vertical, which is the mean point of interaction of protons at this energy.

The second observable which is estimated when the cosmic ray observatories analyze their data is the energy of the primary particle using the signal in the ground detectors, where available. The Pierre Auger Collaboration estimate the energy of the primary particle by using the signal recorded by the ground detectors situated 1000 meters away (S(1000)) from the shower axis \cite{Abraham:2004dt}:
\be
E(EeV) \!=\! 0.12\Big(\!\sqrt{1+11.8(sec\;\theta\!-\!1)^{2}}\;S(1000) \Big)^{1.05}
\label{energy}
\ee
where $\theta$ represents the zenith angle of the incoming primary particle and the signal S(1000) is proportional to the number of charged particles which are recorded by the ground detectors.

\begin{widetext}

\begin{figure}[ht]
\includegraphics[scale=0.92]{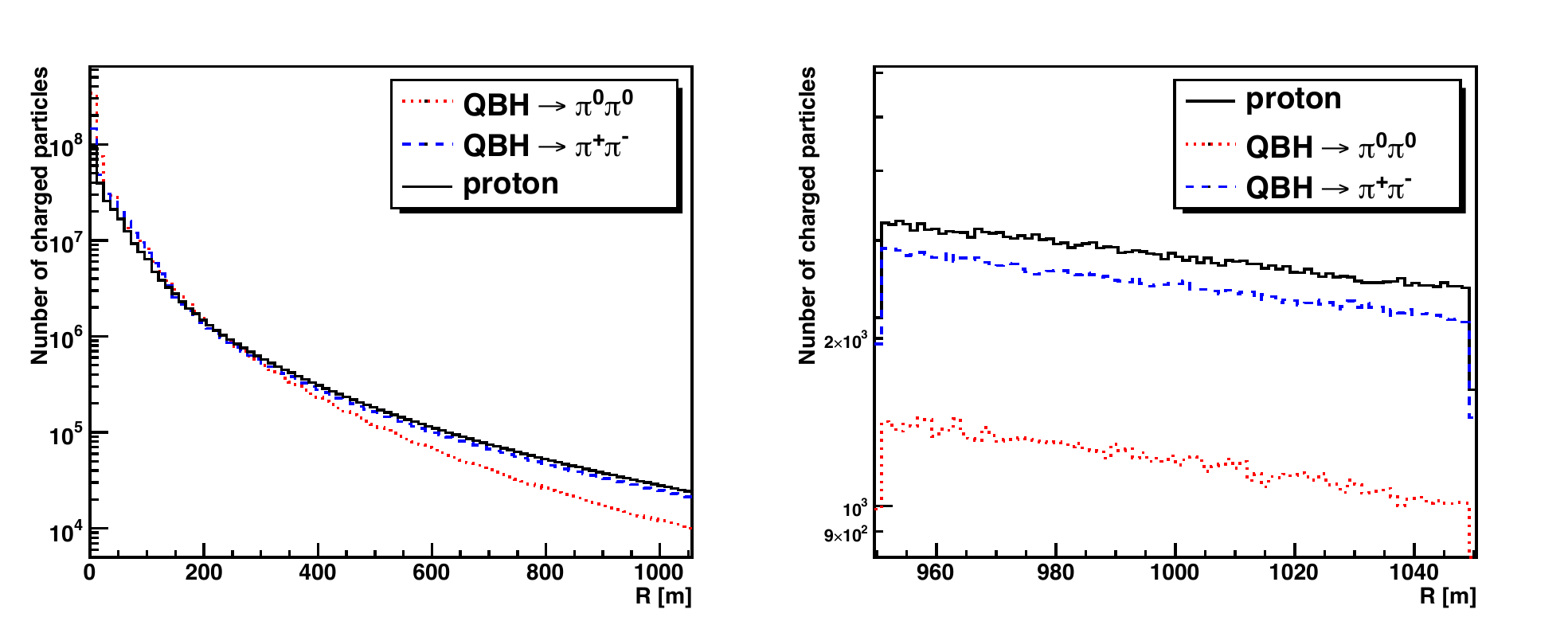}
\caption{ Lateral distribution functions (number of charged particles versus the distance from shower axis) for the three cases: standard model type proton induced showers (black straight line), $QBH\rightarrow\pi^{+}\pi^{-}$ (blue dashed line) and $QBH\rightarrow\pi^{0}\pi^{0}$ (red dashed line). The plot on the left shows the distribution starting from the core of the shower, while the plot on the right represents a zoom in the region around 1000 m. These are the average values calculated over 50 simulations per each case. The number of bins in each plot is 100. }
\label{plot3}
\end{figure}
\end{widetext}

Fig.~\ref{plot3} represents the lateral distribution function of charged particles at the observation level (as a reminder this part of the analysis is pertinent to cosmic ray observatories which can record the particles which arrive on the ground) as a function of the distance from the shower axis. We observe that the density of charged particles is greater for the case of standard model showers in comparison with black hole induced showers.

While the integrals of the curves in Fig.~\ref{plot3} are equal in reality, since the total energies are the same, they do not appear to be so due to saturation of the detectors in the core of the shower where the quantum black hole case dominates the standard model case.

The right panel of Fig.~\ref{plot3} represents a zoom for the lateral distribution functions in the region from 950 to 1050 meters from the shower axis. 
The ratio of the number of charged particles falling in this region in the case of the standard model proton induced showers to the number of particles falling in this region for the two types of quantum black hole decays considered here is: 
\be
\frac{\rho_{ch}^{p}}{\rho_{ch}^{qBH\rightarrow\pi^{+}\pi^{-}}} \simeq 1.15,
\ee
\be
\frac{\rho_{ch}^{p}}{\rho_{ch}^{qBH\rightarrow\pi^{0}\pi^{0}}} \simeq 2.36,
\ee
with $\rho_{ch}^{p}$ representing the number of charged particles for a standard model proton
shower, while $\rho_{ch}^{qBH\rightarrow\pi^{+}\pi^{-}}$ and $\rho_{ch}^{qBH\rightarrow\pi^{0}\pi^{0}}$ represent the number of charged particle for the two quantum black hole decay channels considered. 

   
Remembering that for all the simulations the initial energies were the same, these plots show that the surface detectors will underestimate the energies of the primary particles when the processes occur via intermediary black hole states. For the case of primary particles with energies of $10^{18}$ eV, the energies reconstructed in this way will appear to be $8.6\times 10^{17}$ in the case of quantum black holes which decay into a pair of charged pions and $4.0 \times 10^{17}$ in the case of quantum black holes which decay into a pair of neutral pions.

Putting everything together, one realizes that if quantum black holes are created as intermediary states, the extensive air showers look very different from the typical standard model proton generated showers. The atmospheric depth for which the density of charged particles is maximum
increases, while the energy calculated from the number of charged particles which reach the ground detectors is underestimated by anywhere between 14$\%$ and 60$\%$. 

\begin{figure}[ht]
\includegraphics[scale=0.45]{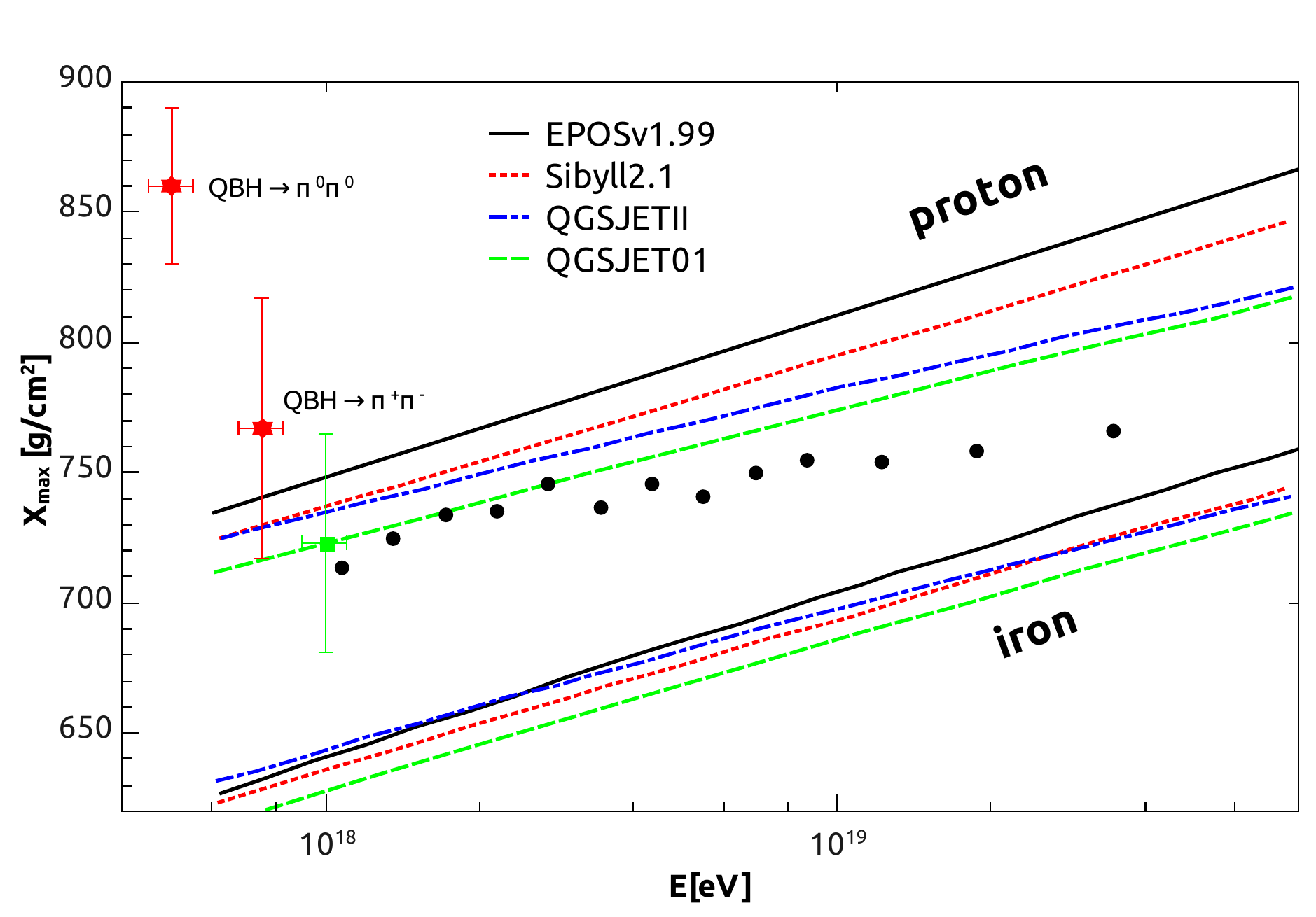}
\caption{Variation of the atmospheric depth for which the density of charged particles is maximum - $X_{max}$ as a function of the energy. The plot presents the Pierre Auger data compared to air shower simulations \cite{Bergmann:2006yz} for different hadronic models \cite{Kalmykov:1993qe, Ostapchenko:2004ss,Pierog:2006qv,Ahn:2009wx}. In addition we include the cases of black holes induced events for the simulations presented above. The energies of the primary particles used as input in our simulations were $10^{18}$ eV. The error bars for our data points represent the RMS of $X_{max}$. The red dots represent the cases in which there is an intermediary quantum black hole, while the green dot represents our numerical simulations for the standard model proton induced showers. The simulations were performed using the QGSJET01 model.}
\label{plot4}
\end{figure}

Fig.~\ref{plot4} shows the variation of $X_{max}$ as a function of the energy. The plot presents the Pierre Auger data
compared to air shower simulations for several hadronic models. It also includes the two data points representing the results of the simulations for extensive air showers produced via back-to-back black hole decays. 

Recapitulating shortly, for our simulations we
considered $10^{18}$ eV protons which produced quantum black holes by interacting with nucleons in the atmosphere. The quantum black
holes decayed back-to-back into pairs of pions (charged or neutral). In 25.8$\%$ of the cases the energies of the 
pions are approximately equal in the reference frame of the experiment, with values on the order of $5\times 10^{17}$ eV. The two pions produce overlapping showers. 
The simulations for these two overlapping showers show $<X_{max}^{QBH\rightarrow\pi^{+}\pi^{-}}>$ = 767 g/cm$^{2}$ (44 g/cm$^2$ larger than for protons when using the same interaction model) for the first decay mode considered and $<X_{max}^{QBH\rightarrow\pi^{0}\pi^{0}}>$ = 860 g/cm$^{2}$ (137 g/cm$^2$ larger than
for protons when using the same interaction model) for the second decay mode. The energies estimated for the primary cosmic rays are $8.6\times 10^{17}$ eV respectively $4.0\times 10^{17}$ eV. These energies need to be compared with the $10^{18}$ eV benchmark
energy, which is the energy one estimates for the protons by performing the same analysis. 

The systematic errors when estimating the
energy are around 22\%. The standard deviations of the $X_{max}$ value are $42$ g/cm$^2$ for the standard model proton induced showers,
 $50$ g/cm$^2$ for the $QBH\rightarrow\pi^{+}\pi^{-}$ case and $30$ g/cm$^2$ for the $QBH\rightarrow\pi^{0}\pi^{0}$ case. The error bars are also represented in Fig.~\ref{plot4}. Note that these values are obtained when considering the first interaction point/starting point at a height of 20 km verticaly.

Only simulations for the
primary particles having energies of  $10^{18}$ eV were performed due to computer power limitations. A simulation of this type takes on the order of a week and the time scale increases with the energy of the primary
particle due to the much larger number of particles produced in the showers. Given the steady increase of $X_{max}$ with the energy in
the numerical simulations shown in Fig.~\ref{plot4} we have strong reason to believe that the same behaviour will be present when
performing numerical simulations for quantum black hole induced showers. Even so, we will not rely on this assumption but perform
simulations at higher energies, but this will be a lengthy process and the findings will be presented in a subsequent letter.  

Having these differences in $<X_{max}>$ and taking into account the number of quantum black holes which are estimated in section III, we further analyse if this small number of quantum black hole events can be separated from the proton induced standard model showers using the $X_{max}$ distributions. 
In order to do that, we considered the flux of protons with energies above $10^{18}eV$ (which we claim to be the background) extrapolated from  \cite{2008PhRvL.101f1101A} with $<X_{max}^{p}>$ = 723 g/cm$^{2}$ and RMS = 55 g/cm$^{2}$ and the flux of quantum black holes (signal). The number of events is multiplied with a factor of 20 to obtain the statistics for 20 years and sprayed into Gau$\ss$ian distributions with the associated $<X_{max}>$ and RMS values, as plotted in Fig.~\ref{plot5} and Fig.~\ref{plot6}. Note that we took into account only the fraction of events seen by the fluorescence detectors from the Pierre Auger Observatory which has a duty cycle of about $13-14\%$.

We calculate the statistical significance $s/ \sqrt{s+b}$ (where {\it s} stands for signal and {\it b} stands for background) in the interval of $X_{max} $[800 - 1100 g/cm$^{2}$]. We find it to be 0.07, 0.59, 3.23, 4.39, 5.35, 6.78 for n = 0, 1, 4, 5, 6 and 7 extra dimensions in the case $QBH\rightarrow\pi^{+}\pi^{-}$, respectively 0.20, 1.53, 8.97, 11.90, 14.83, 18.03 for n = 0, 1, 4, 5, 6 and 7 extra dimensions in the case $QBH\rightarrow\pi^{0}\pi^{0}$.
From Fig.~\ref{plot6} we obtained the number of quantum black holes which has an $X_{max}$ greater than $990 g/cm^{2}$ which are totally separated by primary protons.
For n = 4, 5, 6 and 7 number of extra dimensions we expect that the Pierre Auger Observatory to record 41, 63, 80, 94 quantum black hole events in 20 years of observation.

\begin{widetext}

\begin{figure}[ht]
\includegraphics[scale=0.9]{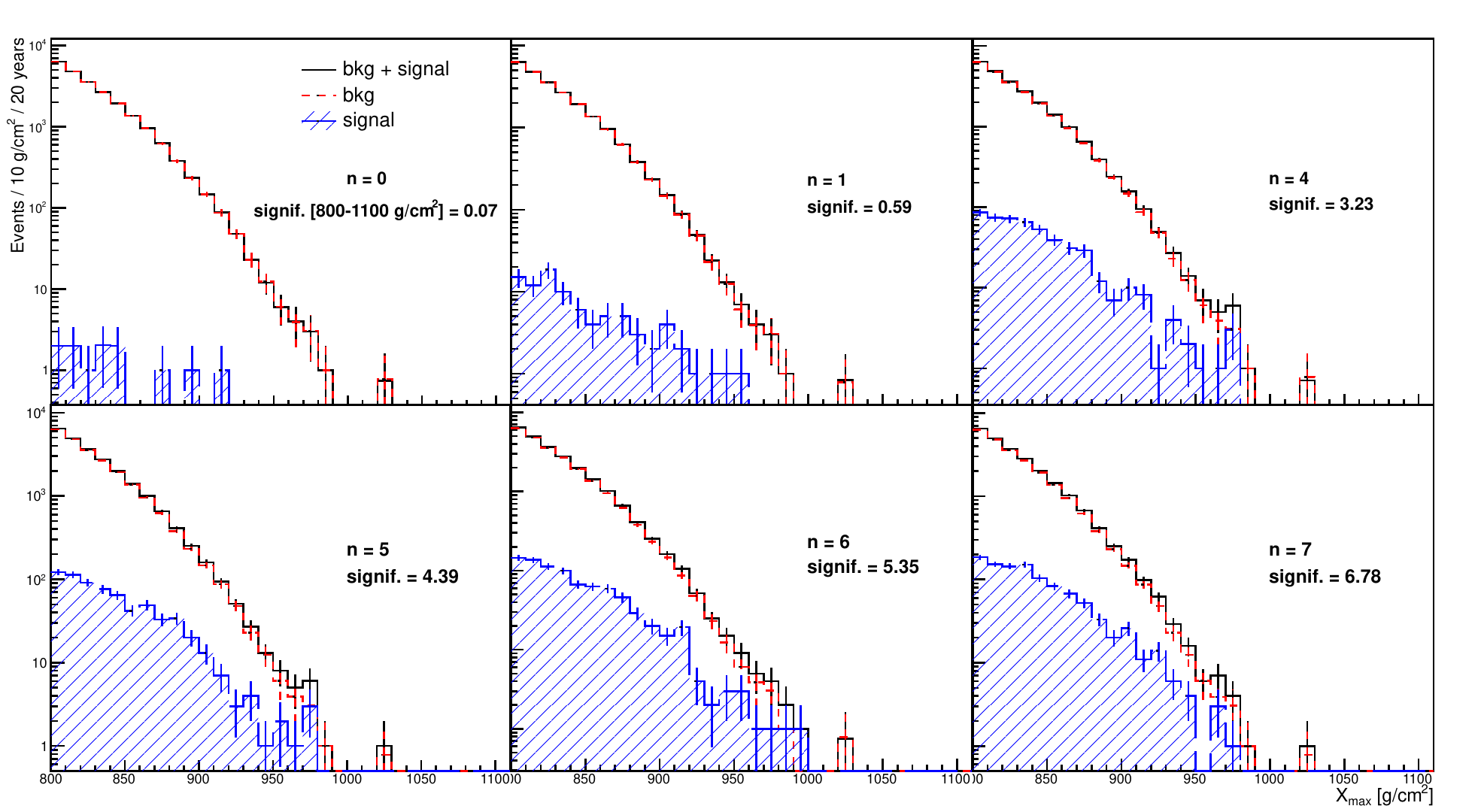}
\caption{Signal (events which produced black holes) and background (protons which interact without producing black holes) estimates for the black hole events visible to the fluorescence detectors produced by 
$10^{18}\pm 10\%$ eV protons in 20 years of statistics at the Pierre Auger Observatory for the $QBH\rightarrow\pi^{+}\pi^{-}$ case. The number of events for the background distribution is 232400, while for the signal we have 40, 274, 1642, 2196, 2776, 3375 events (for n = 0, 1, 4, 5, 6, 7 extra dimensions). The blue filled lines represent the signal distribution, the red 
dashed lines represent the background and the black straight lines represent the signal plus background distribution. The Planck mass was assumed to be 5 TeV. 
The statistical significance is calculated for the range [800-1100 g/cm$^{2}$],
and $n$ represents the number of extra dimensions. We took into account the systematic errors for the $X_{max}$ reconstruction at the Pierre Auger Observatory (20 g/cm$^{2}$). The error bars represent the statistical 
uncertainties.}
\label{plot5}
\end{figure}

\begin{figure}[ht]
\includegraphics[scale=0.9]{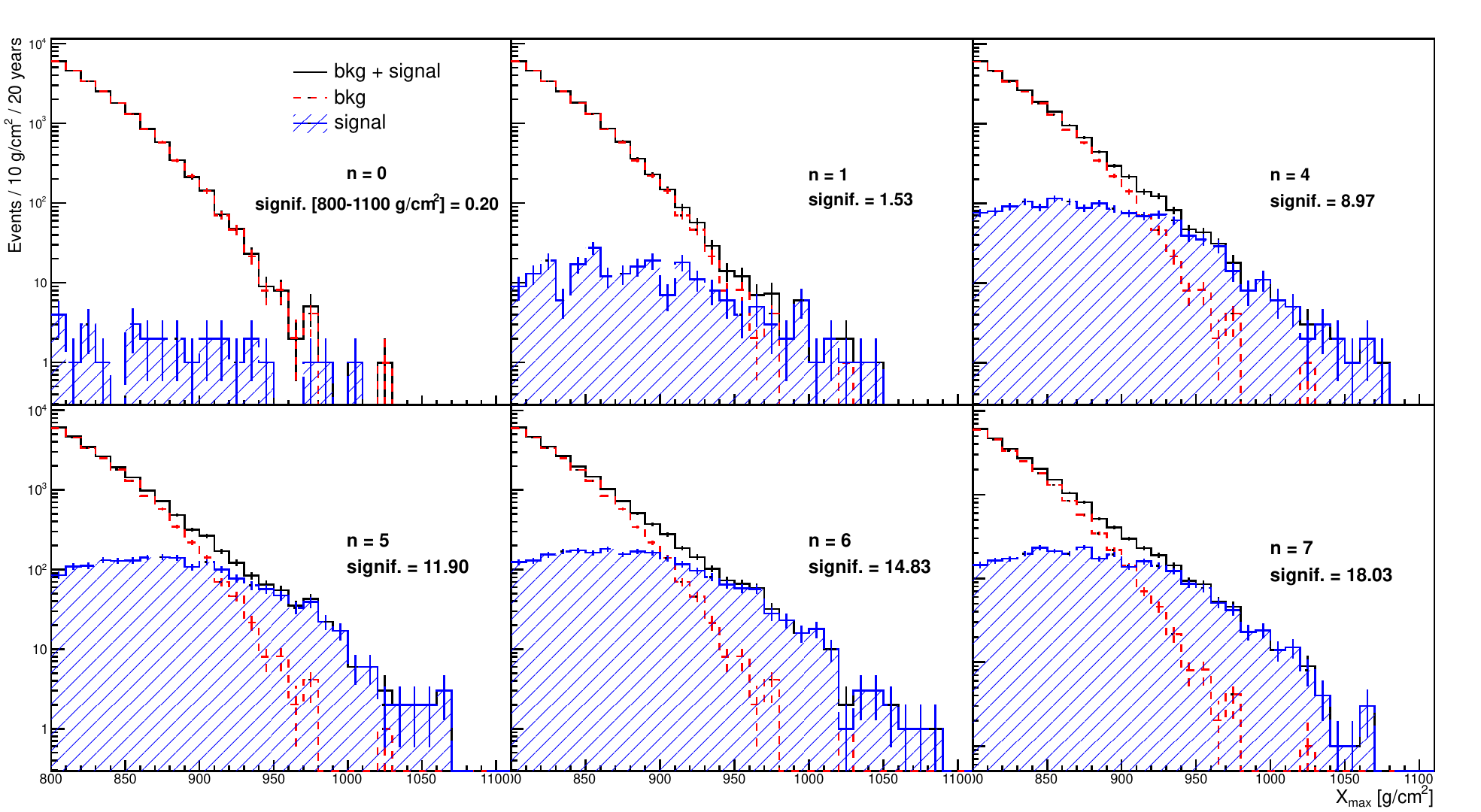}
\caption{Signal (events which produced black holes) and background (protons which interact without producing black holes) estimates for the black hole events visible to the fluorescence detectors produced by 
$10^{18}\pm 10\%$ eV protons in 20 years of statistics at the Pierre Auger Observatory for the $QBH\rightarrow\pi^{0}\pi^{0}$ case. The number of events for the background distribution is 232400, while for the signal we have 40, 274, 1642, 2196, 2776, 3375 events (for n = 0, 1, 4, 5, 6, 7 extra dimensions). The blue filled lines represent the signal distribution, the red 
dashed lines represent the background and the black straight lines represent the signal plus background distribution. The Planck mass was assumed to be 5 TeV. 
The statistical significance is calculated for the range [800-1100 g/cm$^{2}$],
and $n$ represents the number of extra dimensions. We took into account the systematic errors for the $X_{max}$ reconstruction at the Pierre Auger Observatory (20 g/cm$^{2}$). The error bars represent the statistical 
uncertainties.}
\label{plot6}
\end{figure}

\end{widetext}

\section{Conclusions and Overlook}

While present day particle accelerators allow us to test for the Planck scale up to the 10 TeV region, ultra high energy cosmic ray observatories provide a unique opportunity to go one order of magnitude higher in energy. At the same time, the ultra high energy cosmic ray data allows for complementary searches to the ones done at the LHC to be performed. 
The Planck scale can be searched for via non-thermal quantum black hole decay signatures. Above the quantum gravity scale, quantum black holes can be created via the collisions of ultrahigh energy cosmic rays with nucleons from the atmosphere. These holes decay instantaneously preferentially into two particles which produce two overlapping hadronic showers. 

The resulting showers have different profiles and $X_{max}$ values when compared
with similar showers generated via purely standard model processes (without intermediary quantum black hole states). The shift in $X_{max}$, when considering $10^{18}$ eV primary ultra high energy protons, is of approximately 44 g/cm$^{2}$ for showers generated by intermediary black holes which decay into pairs of charged pions and it gets much larger, 137 g/cm$^2$ more exactly, for showers generated by black holes which decay into pairs of neutral pions. On top of this, the primary particle energies estimated using the number of charged particles recorded by detectors situated at roughly 1000 meters from the shower axis are underestimated by  14$\%$ in the first case and by 60$\%$ in the second case.  

The plots in Fig. \ref{plot5} and Fig. \ref{plot6} show that the signal significance is large enough for this signature to be detected at least in several of the cases under consideration. Given the constant improvements on exposure, detector efficiency and on the data analysis part, it is very likely for the situation to get much better in the near future. Our analysis was performed only for the case of the Pierre Auger Observatory, but a similar exercise can be carried out for other present day or future cosmic ray observatories. We emphasise again here that the Telescope Array and the JEM-EUSO experiment are two more viable candidates for performing these searches and we plan to carry out similar analyses for these two experiments in the near future. 

Therefore, we conclude that this signature is very suitable to be used for performing quantum black hole searches in the data recorded by cosmic ray observatories. When discovered above a certain energy, this signature would be a clear indication of the presence of a threshold such as the one due to reaching above the value of the Planck scale. Due to limited computer power, we only produced simulations at $10^{18}$ eV with the initial particles oriented vertically towards the Earth and with the primary interaction point at an altitude of 20 km were performed so far. Our results are so far qualitative and it is in our future plan to perform more numerical simulations for other possible decay channels over a broader range of energies and considering primary particles striking the atmosphere at different angles with respect to the ground. 

\vspace{5mm}
\paragraph*{Acknowledgements:} We wish to thank Prof. Dr. Octavian Sima of the University of Bucharest for useful discussions. This work was supported in part by the European Cooperation in Science and Technology (COST) Action MP0905 ``Black Holes in a Violent  Universe". N.A., L.I.C. and O.M. were supported by research grant: UEFISCDI project PN-II-RU-TE-2011-3-0184.

\bibliography{theorybib}

\end{document}